\documentclass[%
aip, adv,%
 amsmath,amssymb,
preprint,%
]{revtex4-1}

\usepackage{graphicx}
\usepackage{dcolumn}
\usepackage{bm}

\begin{document}

\preprint{Preprint }

\title{Steric effect of water molecule clusters on electrostatic interaction and electroosmotic transport in aqueous electrolytes: a mean-field approach}

\author{Jun-Sik Sin}
 \email{js.sin@ryongnamsan.edu.kp \\}
\affiliation{
  Department of Physics,\\
 {\bf Kim Il Sung} University,\\
  Pyongyang 999093, Democratic People's Republic of Korea 
}
\affiliation{%
  Natural Science Center,\\
 {\bf Kim Il Sung} University,\\
  Pyongyang 999093, Democratic People's Republic of Korea 
}%
\author{Yong-Man Jang}
\affiliation{%
  Natural Science Center,\\
 {\bf Kim Il Sung} University,\\
  Pyongyang 999093, Democratic People's Republic of Korea 
}%
\author{Chol-Ho Kim}
\affiliation{%
  Natural Science Center,\\
 {\bf Kim Il Sung} University,\\
  Pyongyang 999093, Democratic People's Republic of Korea 
}%
\author{Hyon-Chol Kim}
\affiliation{%
  Natural Science Center,\\
 {\bf Kim Il Sung} University,\\
  Pyongyang 999093, Democratic People's Republic of Korea 
}

\begin{abstract}
We theoretically study the size effect of water molecule clusters not only on electrostatic interaction between two charged surfaces in an aqueous electrolyte but also on electroosmotic transport in a nanofluidic channel. Applying a free energy based mean-field approach accounting for different sizes of ions and water molecule clusters, we derive a set of coupled equations to compute electrostatic and electroosmotic properties between charged surfaces. We verify that the smaller the size of a water cluster, the stronger the electroosmotic transport in nanofluidic channels. In addition, we find that an increase in size of a water cluster yields a decrease in electrostatic interaction strength between similar or oppositely charged planar surfaces.
\end{abstract}

\maketitle
Everybody will arguably say that water is the most important substance in human life.
It is justified by noting that in all living organisms, water plays the role of a solvent as well  as a medium for solute-solute interactions and therefore is involved in important physical and chemical processes such as transport of substances through cell membranes and interaction between charged objects \cite%
{Israel_book_2011, Lyklema_book_2005, Brov_2008}.

It is a common sense that natural water exists as a dynamic and flexible water cluster containing 14$\sim$16 molecules linked by hydrogen bonds, while small molecules water exists as a water cluster which contains 5$\sim$7 molecules. It is also known that hydrogen bonded network structure lowers diffusivities and slows the biological and chemical processes.
In particular, small molecules water (also referred to as small cluster water) having fewer hydrogen bonds can easily pass through living cell membranes \cite%
{Shulta_2012, Symons_2001}, and has been thereby used for promoting human health \cite%
{ Lo_2000, Liu_1996} and treating human diseases \cite%
{Yang_2018}.

However, most of theoretical studies of small molecules water have been mainly focused on electronic structural properties of water clusters and quantum chemical interaction between them - not on size effect of water molecule clusters \cite%
{Peet_1995, Gregory_1997, Shields_2010, Zhu_2014, Tobias_2016}.

In fact, the Poisson-Boltzmann approach \cite%
{Gouy_JPhysF_1910, Chapman_PhilosMag_1913}, which has been widely used in order to represent electrolyte solutions is a mean-field approach which assumes point-like ions and constant permittivity, and neglects statistical correlations. For low electrostatic potentials, the Poisson-Boltzmann approach has successfully determined ionic concentration profiles close to surfaces. Unfortunately, the Poisson-Boltzmann theory strongly overestimates ionic concentrations close to charged surfaces for highly charged surfaces and multivalent ions.

Many researchers have been proposed various methods to include the steric effect of ions and water molecules into the Poisson-Boltzmann approach \cite%
{Bikerman_PhilosMag_1942, Wicke_ZEC_1952, Iglic_JPhys_1996, Andelman_PRL_1997, Chu_BJ_2007, Kornyshev_JPCB_2007, Li_PRE_2012, Boschitsch_JCC_2012, Iglic_EA_2015, Sin_EA_2016, Iglic_GP_2017}. However, to the best of our knowledge, there is no mean-field approach based study of steric effect of water molecule clusters.

In this communication, we extend the previous mean-field methods \cite%
{Boschitsch_JCC_2012, Iglic_EA_2015, Sin_EA_2016} to the case of electrolytes consisting of water molecule clusters. The mean-field approach clearly shows how size of a water molecule cluster affects electroosmotic transport in nanochannel or electrostatic interaction between similar or oppositely charged surfaces. For the sake of clarity, we consider a symmetric electrolyte containing negative ions of charge $-ze_0$ and positive ions of charge $ze_0$, where $e_0$ is the electron charge. We assume that the elementary unit of water being solvent of the electrolyte is a water cluster containing $N_{wc}$ water molecules.  The bulk ion number density of the negative and positive ions is $n_b$. For the case of point-like ions, our approach is reduced to the original Poisson-Boltzmann equation \cite%
{Gouy_JPhysF_1910, Chapman_PhilosMag_1913}.

Within mean-field approximation, the total free energy, $F$, can be written in terms of the local electrostatic potential and the ionic concentrations. 

We consider the size effect of water molecule clusters on the electrostatic interaction between two planar surfaces (similar or oppositely charged) separated by a distance $W_h$ in an aqueous electrolyte. The transverse direction is denoted by $r$; the bottom plate is placed at $r=-H$ and the top plate $r=H$,i.e. $W_h=2H$. The resulting electrostatic properties between the plates and the electrolyte solution should be addressed by setting the total free energy as follows \cite%
{Bikerman_PhilosMag_1942, Wicke_ZEC_1952, Iglic_JPhys_1996, Andelman_PRL_1997,  Chu_BJ_2007, Kornyshev_JPCB_2007, Li_PRE_2012, Boschitsch_JCC_2012, Iglic_EA_2015, Sin_EA_2016, Iglic_GP_2017} :
\begin{equation}
F = \int d {\bf{r}}\left[ { - \frac{{\varepsilon _0 \varepsilon_r |\nabla \psi\left(\bf r\right)|}^2}{2}  + e_0 z\psi \left( {\bf{r}} \right)\left( {n_ +   - n_ -  } \right) - \mu _ +  n_ +   - \mu _ -  n_ -   - \mu _{wc} n_{wc}  - Ts} \right].
\label{eq:1}
\end{equation}
While the local electrostatic potential is denoted by $\psi \left( {\bf{r}} \right)$, the number densities of positive and negative ions and water molecule clusters  are expressed as $n_i \left( {\bf{r}} \right)\left(i=+, -, wc\right) $, respectively.
In Eq. (\ref{eq:1}), the first term describes the self energy of the electrostatic field, where $\varepsilon _0$  and $\varepsilon_r$ stand for the vacuum permittivity and the relative permittivity of bulk aqueous electrolyte, 78.5, respectively.  The second term means the electrostatic energy of the ions. The next three terms are needed to couple the system to a bulk reservoir, where  $
\mu _i \left(i=+, -\right)$ means the chemical potential of the ions and $\mu _{wc}$ corresponds to the chemical potential of water clusters. $T$  and $s$ are the temperature and the entropy density, respectively. 

In this paper, we have focused on small cluster water limited to the magnitudes of surface charge densities smaller or equal than $0.05 C/m^2$, so that we can neglect a decrease in permittivity of electrolyte solution due to orientation ordering of water dipoles \cite%
{Iglic_2014, Marcovitz_2015}. 
It was shown recently by using Monte Carlo simulations \cite%
{Marcovitz_2015} that even for high salt concentrations orientational ordering of water dipoles is increased in directions towards the charged surface (without van der Waals forces) which is in agreement with theoretical predictions \cite%
{Iglic_2014}. In particular, it should be emphasized that near a charged surface,  the assumption that almost all water molecules in electrolyte solution belong to water shells around the ions, while free water molecules are excluded, is against the results of simulations which clearly show the increased water ordering in the direction towards the charged surface even for high salt concentrations \cite%
{Marcovitz_2015}.

The change in size of water clusters due to electric field close to the charged surface is neglected for the sake of simplicity even the electric field near the charged surface is very large.

On the other hand, we know that due to the same reason as in \cite%
{Iglic_JPhys_1996, Boschitsch_JCC_2012, Onuki_PRL_2017}, Eq. (\ref{eq:2}) works very well when we deal with low bulk ionic concentration and low zeta potential:
\begin{equation}
1 = n_ +  V_ +  + n_ -  V_ - + n_{wc} V_{wc}, 
\label{eq:2}
\end{equation}
where $V_{+}$, $V_{-}$ and $V_{wc}$ are the effective volumes of a positive ion, a negative ion and a water molecule cluster, respectively.

The Lagrangian of the electrolyte with an undetermined multiplier can be expressed by considering the constraint condition.
\begin{equation}
L = F - \int {\alpha \left( {\bf{r}} \right)} \left( {1 - n_ +  V_ +   - n_ -  V_ -   - n_{wc} V_{wc} } \right)d{\bf{r}},
\label{eq:3}
\end{equation}
where $\alpha$  is a local Lagrange parameter. 

For low ionic concentrations, it is noticeable that entropy density can be calculated by Eqs.(\ref{eq:4})  and  (\ref{eq:5}) as in \cite%
{ Boschitsch_JCC_2012}:
\begin{equation}
s =   k_B \ln w,
\label{eq:4}
\end{equation}
where $k_B$ is the Boltzmann constant and $w$ is the number of configurations in a unit volume of the electrolyte
\begin{equation}
w = \frac{{(n_{wc}  + n_ +   + n_ -  )!}}{{n_{wc} !n_ +  !n_ -  !}}.
\label{eq:5}
\end{equation}

From planar symmetry of the charged surfaces, the present study is reduced to an one-dimensional problem.

Minimizing the corresponding free energy with respect to  $n_{i}\left(r\right)\left(i=+,-,wc\right)$ and $\psi\left(r\right)$, we derive the self-consistent equations determining $n_{i}\left(r\right)$ and $\psi\left(r\right)$.
In the same way as in \cite%
{ Boschitsch_JCC_2012}, the number densities of ions and water molecule clusters can be obtained 
\begin{equation}
n_ +   = \frac{{n_{b} \exp \left( { - V_ +  h - e_0 z\beta \psi } \right)}}{D},
\label{eq:6}
\end{equation}
\begin{equation}
n_ -   = \frac{{n_{b} \exp \left( { - V_ -  h + e_0 z\beta \psi } \right)}}{D},
\label{eq:7}
\end{equation}
\begin{equation}
n_ {wc}   = \frac{{n_{ wcb} \exp \left( { - V_ {wc}  h } \right)}}{D},
\label{eq:8}
\end{equation}
where $\beta=1/\left(k_{B}T\right)$, $D = n_{b} V_ +  \exp \left( { - V_ +  h - e_0 z\beta \psi } \right) + n_{b} V_ -  \exp \left( { - V_ -  h + e_0 z\beta \psi } \right) + n_{wc} V_{wc}$, $h\left(r\right)=\alpha\left(r\right)-\alpha\left(r=\infty\right)$ and $n_{wcb}$ denotes the bulk number density of water molecule clusters.

In addition to the above ones, another equation is derived by combining Eqs. (\ref{eq:2}), (\ref{eq:6}), (\ref{eq:7}) and (\ref{eq:8}):
\begin{equation}
n_{b} \left( {e^{ - V_ +  h - \beta ze_0\psi }  - 1} \right) + n_{ b} \left( {e^{ - V_ -  h + \beta ze_0\psi }  - 1} \right) + n_{wcb} \left( {e^{ - V_{wc} h}  - 1} \right) = 0.
\label{eq:9}
\end{equation}
Euler-Lagrange equation for $\psi \left( {r} \right)$  yields the Poisson equation:
\begin{equation}
  \varepsilon _0 \varepsilon _r \frac{d^2 \psi }{dr^2} =  - e_0 z\left( {n_ +   - n_ -  } \right).
\label{eq:10}
\end{equation}
When we consider the case of $N_{wc}$=1, our theory is reduced to one of \cite%
{Boschitsch_JCC_2012}.

As proved in \cite%
{Andelman_JPCB_2009}, osmotic pressure between two charged surfaces can be derived from the following expression 
\begin{equation}
f - \left( {\partial f/\partial \left(\nabla \psi \right)} \right)\nabla \psi = constant =  - P,
\label{eq:14}
\end{equation}
where the constant is the negative of the local pressure $P$ that is defined as the sum of the osmotic pressure ($\Pi$) and the bulk pressure ( $P_{bulk}$), i.e.,$ P= \Pi +P_{bulk}$.
Combining Eq. (\ref{eq:1}) and Eq. (\ref{eq:14}), the following relation is obtained: 
\begin{equation}
\begin{array}{l}
 \Pi  =  - \frac{{\varepsilon _0 \varepsilon_r |\nabla \psi|^2}}{2}  + k_B Th.
\label{eq:15}
 \end{array}
\end{equation}
To derive the above formular we follow the corresponding one of \cite%
{Sin_JCP_2017}.
Here we introduce the following dimensionless variables for more convenient determination of physical quantities:
\begin{equation}
\begin{array}{l}
 \bar r = r/H,\bar \psi  = e_0 z\beta \psi, \lambda  = \sqrt {\frac{{\varepsilon _0 \varepsilon _r k_B T}}{{2n_b e_0^2 z^2 }}}, \bar \lambda  = \lambda /H,\bar h = hV_{wc}, \\ 
 \bar V_{i} = V_{i}/V_{wc} \left(i=+,-\right),\theta  = \frac{1}{{n_{wcb} V_{wc} }},\bar n_b  = \frac{{n_b }}{{n_{wcb} }}. 
 \end{array}
\label{eq:11}
\end{equation}
The governing equations are transformed into the following equations.
\begin{equation}
\frac{{d^2 \bar \psi }}{{d\bar r^2 }} = \frac{\theta }{{2\bar \lambda ^2 }}\frac{{\exp \left( {\bar \psi  - \bar V_ -  \bar h} \right) - \exp \left( { - \bar \psi  - \bar V_ +  \bar h} \right)}}{{\exp \left( { - \bar h} \right) + \bar n_b \bar V_ -  \exp \left( {\bar \psi  - \bar V_ -  \bar h} \right) + \bar n_b \bar V_ +  \exp \left( { - \bar \psi  - \bar V_ +  \bar h} \right)}}
\label{eq:12}
\end{equation}
\begin{equation}
\frac{{d\bar h}}{{d\bar r}} = \frac{{\bar n_b \left( {\exp \left( {\bar \psi  - \bar V_ -  \bar h} \right) - \exp \left( { - \bar \psi  - \bar V_ +  \bar h} \right)} \right)\frac{{d\bar \psi }}{{d\bar r}}}}{{\exp \left( { - \bar h} \right) + \bar n_b \bar V_ -  \exp \left( {\bar \psi  - \bar V_ -  \bar h} \right) + \bar n_b \bar V_ +  \exp \left( { - \bar \psi  - \bar V_ +  \bar h} \right)}}
\label{eq:13}
\end{equation}
In fact, the expression for osmotic pressure Eq. (\ref{eq:15}) can be also obtained by calculating the first integral of Eq. (\ref{eq:12}) \cite%
{Evans_1999, Iglic_2014}. 
First, Eq. (\ref{eq:12}) is multiplied by $d \bar \psi/ d\bar r$ and considered Eq. (\ref{eq:13}) to get the following equation:
\begin{equation}
\frac{{d^2 \bar \psi }}{{d\bar r^2 }}\frac{d\bar \psi}{d\bar r}= \frac{\theta }{{2\bar \lambda ^2 \bar n_b}}\frac{{d\bar h}}{{d\bar r}} 
\label{eq:135}
\end{equation}
As a result, integrating Eq. (\ref{eq:135}) with use of Eq. (\ref{eq:11}) yields the same equation as Eq. (\ref{eq:15}).
We can easily know that in case of $V_{wc}=V_+=V_-$, Eq. (\ref{eq:15}) is reduced to the corresponding expression obtained in \cite%
{Iglic_2014}.

Next, we consider the electroosmotic transport in a slit-like nanochannel to probe how biological and physicochemical processes in living organisms get fast by small molecules water. 
In fact, in order to represent the ion transport through a cell membrane, the authors of \cite%
{Olga_2016} addressed the electroosmotic transport in a nanochannel between two semipermeable membranes.
 We can know that electrostatics for the geometry contains the corresponding one for the case when considering electrostatic interaction between similarly charged surfaces.  
We imagine that an external axial electric field is applied to the nanochannel in order to trigger a steady, one dimensional, and fully developed  electroosmotic flow with a little Reynolds number.
The equation governing electroosmotic transport can be simply expressed as:
\begin{equation}
\eta \frac{d^2 u}{dr^2} + e_0 z\left( {n_ +   - n_ -  } \right)E_{\bot} = 0
\label{eq:16}
\end{equation}
with  $\eta$ being the dynamic viscosity of water and $E_{\bot}$ being the employed constant axial electric field. $u$ means electroosmotic velocity resulting from the electrostatic potential profile between two charged surfaces.
Using Eqs. (\ref{eq:6}) and (\ref{eq:7}), we can finally express Eq. (\ref{eq:17}) in dimensionless form as
\begin{equation}
\frac{d^2\bar u}{d\bar r^2} - \bar E_{\bot}\frac{\theta }{2{\bar \lambda ^2 }}\frac{{\exp\left({\bar \psi } -\bar V_ -  \bar h \right)-\exp \left( -\bar \psi - \bar V_ +  \bar h\right)}}{{\exp \left( { - \bar h} \right) + \bar n_b \bar V_ -  \exp \left( {\bar \psi  - \bar V_ -  \bar h} \right) + \bar n_b \bar V_ +  \exp \left( { - \bar \psi  - \bar V_ +  \bar h} \right)}} = 0,
\label{eq:17}
\end{equation}
where $\bar E_{\bot} = E_{\bot}/E_0$($ E_0  = k_B T/\left( ze_0 H \right)$ is the electric field caused by the potential $k_{B} T/ze_0$ operating over half nanochannel height $)$  and $\bar u = u/u_0$( $u_0  = \left( {\varepsilon _0 \varepsilon _r } \right)\left( {k_B T/ze_0} \right)E_0 /\eta$ $)$ is the velocity scale considered equal to the Helmholtz - Smoluchowski velocity that results from a system zeta potential of magnitude $k_{B}T/ze_0$ in the presence of an electric field $E_0$. 

Using the corresponding electrostatic potential $\bar \psi$ obtained from Eq. (\ref{eq:12}) and (\ref{eq:13}), $\bar u$  will be obtained by numerically solving Eq. (\ref{eq:19}) in the presence of the following wall-boundary conditions:
\begin{equation}
\left( {\bar u_{} } \right)_{\bar r =  \pm 1}  = 0,\left( \frac{d{\bar u }}{d\bar r} \right)_{\bar r = 0}  = 0
\label{eq:18}
\end{equation}
For the clarity, we assume that the plane  ($\bar r = \pm1$) having the zeta potential ($\bar \zeta$) is the plane of no slip for the axial flow velocity, disregarding other effects such as changes in the permittivity and the viscosity in the stern layer.
Combining Eq. ({\ref{eq:10}), Eq. ({\ref{eq:16}}) and Eq. ({\ref{eq:18}}) yields the relationship between dimensionless electroosmotic velocity and dimensionless electrostatic potential, as in \cite%
{Das_PRE_2008}.
\begin{equation}
\bar u =  - \bar E_{\bot}\left( {\bar \zeta   - {{\bar \psi }}}\right)
\label{eq:19}
\end{equation}

Eq. (\ref{eq:19}) indicates that the electroosmotic velocity linearly varies with the electrostatic potential, and the magnitude of the electroosmotic velocity is proportional to the difference between electrostatic potential and zeta potential.
We take the midpoint between the two charged surfaces as the origin of coordinates and choose the positive r-axis to point top.
Electrostatic potential and number densities of ions and water molecule clusters are simultaneously obtained from the numerical solution of the set of coupled equations Eqs. (\ref{eq:6})-(\ref{eq:10}) as a function of position in between two charged surfaces or walls.  Due to the same reason as in \cite%
{Das_PRE_2011}, we have used the bulk concentrations of negative and positive ions as the concentration of ions in the reservoirs connected to the nanochannel.

 For all the calculations, the temperature and the bulk ionic concentration are $T=300K$ and $c_b=0.01M$ (i.e. $n_b=$ 0.01 $\times$ Na $\times$ 1000, where Na is Avogadro number) while the size of a positive ion is $V_{+}=0.03nm^3$.
The magnitude of surface charge density is chosen as $|\sigma|=0.05C/m^2$, in the range of typical values for biological membrane surface, like in \cite%
{McLau_1989, Cevc_1990, Dobrzynska_2013, Sinha_2017}.
\begin{figure}
\begin{center}
\includegraphics[width=1\textwidth]{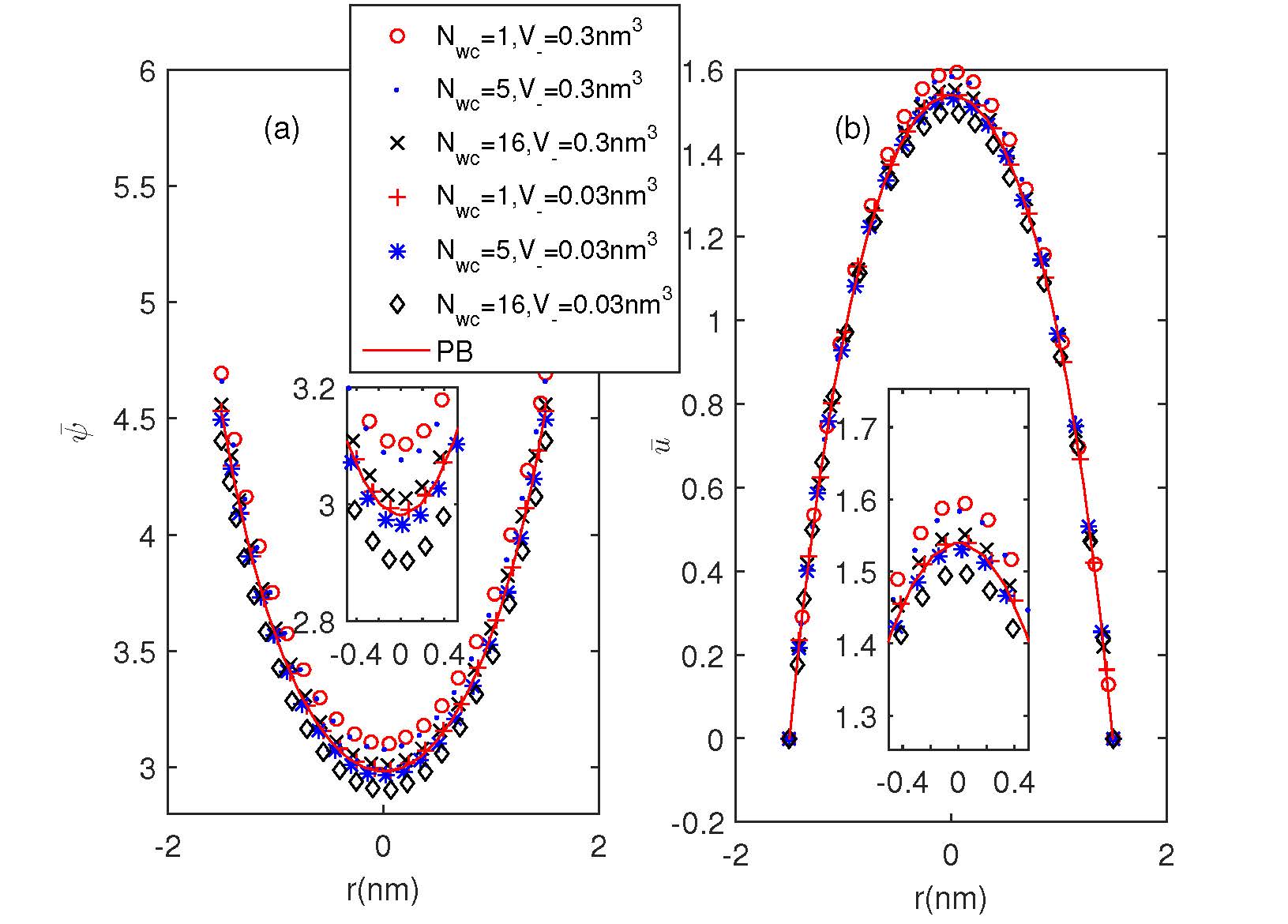}
\caption{(Color online) The dimensionless electrostatic potential profiles (a), and  dimensionless electroosmotic velocity profiles (b)  in a nanochannel having $\sigma\left(x=H\right)=\sigma\left(x=-H\right)=+0.05C/m^2$ for different ion sizes. $\bar E_{\bot}=1, H=1.5nm$.}
\label{fig:1}
\end{center}
\end{figure}
\begin{figure}
\begin{center}
\includegraphics[width=1\textwidth]{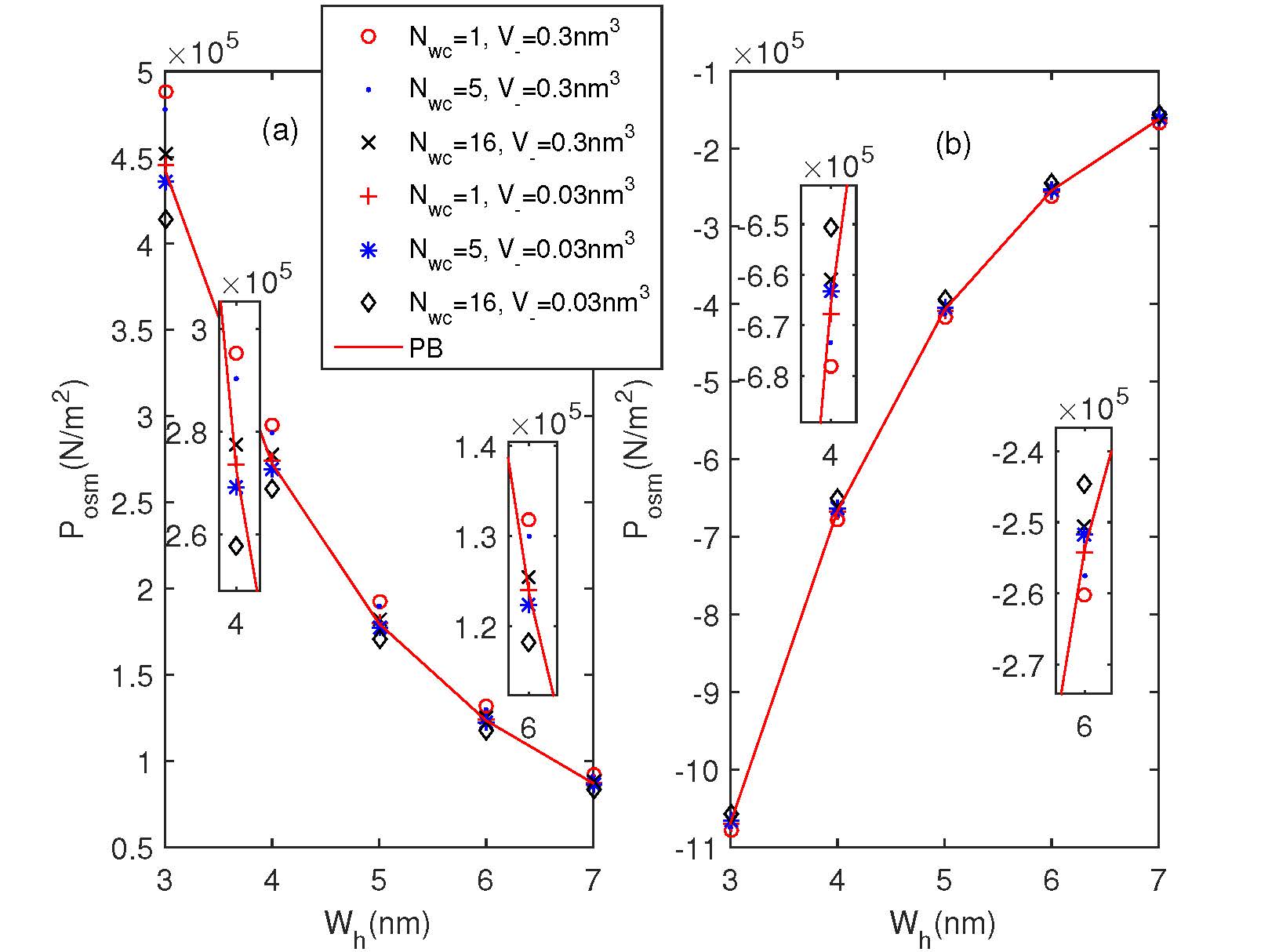}
\caption{(Color online) Osmotic pressure as a function of the separation distance  between similarly charged surfaces for $\sigma\left(x=H\right)=\sigma\left(x=-H\right)=+0.05C/m^2$ (a)  and between oppositely charged surfaces for $\sigma\left(x=H\right)=-\sigma\left(x=-H\right)=+0.05C/m^2$ (b) for different ion sizes. Other parameters and  the meanings of symbols are the same as  in Fig. {\ref{fig:1}}.}
\label{fig:2}
\end{center}
\end{figure}
In Fig. \ref{fig:1}(a), we show the electrostatic potential profiles for different sizes of a water cluster by using our approach. For comparison,  the result of original Poisson-Boltzmann approach is also presented.
The key finding is that for the case where finite ion size is considered, increasing the size of a water molecule cluster $N_{wc}$ results in the lowering of the electrostatic potential across the entire nanochannel. This contrasts with the original Poisson-Boltzmann scheme which leads to only one electrostatic profile regardless of the size of a water molecule cluster. 
In fact, as can be seen in Eq. (\ref{eq:5}), entropy density decreases with increasing the size of a water molecule cluster because a large size of a water molecule cluster means a smaller number of water clusters. As a consequence, the entropy contribution to the free energy becomes small (i.e. thermal motion of ions and water clusters becomes weaker) and in turn counterions in the electrolyte solution can be more easily accumulated to the charged surface. Therefore, electrostatic potential for a small size of water cluster is larger than corresponding ones for a larger size of water cluster.

 As can be seen in Fig. {\ref{fig:1}(b), a large size of a water molecule cluster produces a lower electroosmotic velocity than that of a smaller size of a water molecule cluster, across the entire nanochannel. This phenomenon is explained by the following facts. First, at the charged surfaces ($\bar r=\pm 1$), the difference in electrostatic potential for different sizes of a water molecule cluster is larger than corresponding one at any position across the entire nanochannel. Then, Eq. (\ref{eq:19}) indicates that the electroosmotic velocity at any position across the nanochannel is proportional to the difference between zeta potential and the electrostatic potential at the position. From the above two facts,  we can assert that increasing the size of a water molecule cluster yields a decrease in electroosmotic velocity.
 The experimental results \cite%
{Shulta_2012, Symons_2001, Lo_2000, Liu_1996} established that small molecules water moves in and out of cell fast compared to natural water. Fig. 1(b) demonstrates that small molecules water allows electroosmotic velocity to increase, and consequently justifies the experimental results.

The electrostatic interaction between similar or oppositely charged surfaces can be calculated using Eq. (\ref{eq:15}).
Fig. {\ref{fig:2}(a) and Fig. {\ref{fig:2}(b) depict the osmotic pressure as a function of the distance between two charged surfaces, for two different ion sizes, $V_{-}$. The result of Poisson-Boltzmann approach is also shown for comparison.
Furthermore, as can be seen from Fig. {\ref{fig:2}}(a) and (b), the osmotic pressure at the small distance separation depends strongly on the size of a water molecule cluster, $N_{wc}$. For the Poisson-Boltzmann approach, this interaction doesn't depend on the size of water cluster. On the other hand,  Fig. {\ref{fig:2}}(a) and (b) demonstrates that finite size of ions also changes the steric effect of water clusters since ions prevents water molecule clusters from occupying the vicinity of the charged walls. 

In conclusion, we have used the previous developed mean-field model \cite%
{Boschitsch_JCC_2012, Iglic_EA_2015, Sin_EA_2016} of electric double layer for description of the effect of the size of water clusters on electrostatic interaction between two charged surfaces and electroosmotic transport in a nanochannel. As a result, we have demonstrated that  increasing the size of a water molecule cluster produces not only a decrease in the strength of electrostatic interaction between similar or oppositely charged surfaces but also a decrease in electroosmotic velocity.
This fact justifies a beneficial effect of small cluster water on electrolyte transport through a narrow confined geometry and interaction between charged objects relevant to fundamental physiological processes. Our results are consistent with the experiments  \cite%
{Shulta_2012, Symons_2001} involving electrolyte solutions in living organisms. 

Since the present study treats surface charge densities smaller or equal than $0.05 C/m^2$, we can neglect change in dielectric permittivity due to the orientational ordering of water dipoles. However, in the case of a middle or high surface charge density, one should consider the infuence of orientational ordering of water dipoles on dielectric permittivity \cite%
{Marcovitz_2015, Iglic_2014} and change in size distribution of water clusters.
It should be emphasized that the electrostatic equations were numerically solved in planar geometry and not in cylindrical geometry. Therefore, the present approach is appropriate only for the tubes or channels with rather large diameters.
In the future, we will further consider for which values of the diameters of the channels our results are applicable, as in \cite%
{Bohinc_2005}.

\end{document}